\documentclass[review]{elsarticle}

\usepackage{bm}
\usepackage[normalem]{ulem}
\journal{Journal of \LaTeX\ Templates}
\usepackage{epsfig}
\usepackage{algorithm}
\usepackage{algorithmicx}
\usepackage{algpseudocode}
\usepackage{amsmath,amssymb}
\usepackage{mathrsfs}
\usepackage{bm}
\usepackage{hyperref}
\usepackage{adjustbox}
\usepackage{multirow} 
\usepackage{subfigure}
\usepackage{graphicx}

\begin{document}

\begin{frontmatter}

\title{Improving Stock Market Prediction via Heterogeneous Information Fusion}

%
%
%


\author[mymainaddress]{Xi Zhang\corref{mycorrespondingauthor}}
\cortext[mycorrespondingauthor]{Corresponding author}
\ead{zhangx@bupt.edu.cn}

\author[mymainaddress]{Yunjia Zhang}
\ead{2011213120@bupt.edu.cn}

\author[mysecondaryaddress]{Senzhang Wang}
\ead{szwang@nuaa.edu.cn}

\author[mymainaddress]{Yuntao Yao}
\ead{yaoyuntao@bupt.edu.cn}

\author[mymainaddress,myfifthaddress]{Binxing Fang}
\ead{fangbx@bupt.edu.cn}

\author[mythirdaddress,myfourthaddress]{Philip S. Yu}
\ead{psyu@uic.edu}

\address[mymainaddress]{Key Laboratory of Trustworthy Distributed Computing and Service, Ministry of Education, Beijing University of Posts and Telecommunications, Beijing 100876, China}
\address[mysecondaryaddress]{College of Computer Science and Technology, Nanjing University of Aeronautics and Astronautics, Nanjing 210016, China}
\address[mythirdaddress]{Department of Computer Science, University of Illinois at Chicago, IL 60607, USA}
\address[myfourthaddress]{Institute for Data Science, Tsinghua University, Beijing 100084, China}
\address[myfifthaddress]{Institute of Electronic and Information Engineering of UESTC in Guangdong, Dongguan Guangdong 523808, China}

\begin{abstract}

Traditional stock market prediction approaches commonly utilize the historical price-related data of the stocks to forecast their future trends. As the Web information grows, recently some works try to explore financial news to improve the prediction. Effective indicators, e.g., the events related to the stocks and the people's sentiments towards the market and stocks, have been proved to play important roles in the stocks' volatility, and are extracted to feed into the prediction models for improving the prediction accuracy. However, a major limitation of previous methods is that the indicators are obtained from only a single source whose reliability might be low, or from several data sources but their interactions and correlations among the multi-sourced data are largely ignored.


In this work, we extract the events from Web news and the users' sentiments from social media, and investigate their joint impacts on the stock price movements via a coupled matrix and tensor factorization framework. Specifically, a tensor is firstly constructed to fuse heterogeneous data and capture the intrinsic relations among the events and the investors' sentiments. Due to the sparsity of the tensor, two auxiliary matrices, the stock quantitative feature matrix and the stock correlation matrix, are constructed and incorporated to assist the tensor decomposition. The intuition behind is that stocks that are highly correlated with each other tend to be affected by the same event. Thus, instead of conducting each stock prediction task separately and independently, we predict multiple correlated stocks simultaneously through their commonalities, which are enabled via sharing the collaboratively factorized low rank matrices between matrices and the tensor. Evaluations on the China A-share stock data and the HK stock data in the year 2015 demonstrate the effectiveness of the proposed model.

\end{abstract}


\end{frontmatter}


\section{Introduction}

Stock market prediction has attracted much attention from academia as well as business. Due to its complexity, to what extent can the stock market be predicted remains an open question.
~The early literature on this problem was based on the Efficient Market Hypothesis (EMH)~\cite{fama1965behavior}, which states that the stock market prices fully reflect all available information. Generally, stock-related information can be roughly categorized into the quantitative data and the qualitative descriptions of the financial standings. Quantitative analysis makes investment decisions based on the publicly available quantitative data (e.g., stock price, standard income, balance sheet, etc)~\cite{PAN201790}. Qualitative analysis, on the other hand, looks at the business itself and tries to make decisions based on the qualitative data such as the management, products and strategies of a company. Both types of analysis are important to develop a successful investment strategy. It is therefore reasonable to obtain necessary and comprehensive information, including both quantitative numbers and qualitative descriptions to predict the future trend of a company stock price. Quantitative data is usually well organized and publicly available through the financial data providers such as Bloomberg and Wind~\footnote{http://www.wind.com.cn/}, and has been successfully applied in technical analysis and quantitative trading. Qualitative information, however, usually lies in the textual descriptions from various data sources including the web media and social media. 
~With the prosperity of Web 2.0, more and more investors are engaged in the web activities to obtain and share the stock-related information in real time. Meanwhile, the posted opinions by the experts and influential people on the stocks can influence other people's decisions due to the rapid propagation of the influence through Internet. The effects are twofold. On one hand, the event information and the users' sentiments on the Web can largely influence the stock price. For example, the false rumor of explosion at White House causes stocks to briefly plunge~\cite{whitehouse}. On the other hand, drastic fluctuations in stock price can lead to the generation and spreading of the relevant information (e.g., viewpoints from authorities), which can in turn affect the public opinions for the future investment strategies. Therefore, it provides researchers unprecedented opportunity to utilize the Web information to facilitate stock analysis.


Considering the high correlations between the stock price and the stock related event information on the Web, event-driven stock prediction techniques aim to extract events from Web news to guide stock investment~\cite{ding2014using,ding2015deep,dingknowledge}. However, their prediction capabilities are limited by the following two challenges. First, the stock related event information collected from Web is very sparse. Although Web news is increasingly available, the amount of events that can be extracted from the Web news is still limited. In addition, the events usually reside in unstructured texts which are difficult to extract. Besides, the same event can be described in different ways by different websites and thus is prone to be identified as different events, leading to increased sparsity. Thus, the similar events yet with different expressions should be merged into one category to reduce the sparsity. Second, there lacks an effective method to analyze the events and quantitively measure their influence on the stock prices. Even if we have successfully extracted the events on a stock, it is still difficult to determine whether the event can cause a positive or negative impact on the stock value. For example, for the event acquisition, Microsoft acquiring LinkedIn leads to the fall of Microsoft's stock price. In contrast, Intel acquiring Altera results in the rise of Intel's stock price. Thus, solely relying on the events to make predictions is not enough. In addition to the events, emotions also play a significant role in decision making. Previous studies in behavioral economics show that financial decisions are significantly driven by emotion and mood. For example, the collective level of optimism or pessimism in society can affect investor decisions \cite{Prechter1999,Nofsinger2005}.
~Due to the recent advances in Natural Language Processing (NLP) techniques, sentiment-driven stock prediction techniques are also proposed by extracting indicators of public mood from social media~\cite{bollen2011twitter,nguyen2015topic,feldman2011stock}, where the positive mood for a stock will probably indicate a rise trend in the price and negative mood will more likely mean a decrease trend. However, relying on the sentiments alone is not sufficient for prediction either. For example, in holidays, people's mood tends to be positive yet it may not really reflect their investment opinions.



To address the above mentioned challenges, we propose to combine both the stock-related events extracted from the Web news and the users' sentiments on social media. For this purpose, effective information integration techniques are required to jointly model their impacts. However, it is extremely difficult to integrate the information from multiple sources which is heterogeneous and interrelated. Specifically, the information may be with different time scales (e.g., hours, days, months) and different structures (e.g., events in news, sentiments on social media). A common strategy in previous studies is to concatenate the features from multiple sources into one compound feature vector.
~However, such a linear predictive model assumes these features from different data sources are independent of each other. In reality, in addition to the linear effects, there are also coupling effects coming from the interactions among multiple sources. For example, a specific event (e.g., breaking the contract) usually results in a specific sentiment (e.g., negative emotion). In addition, even within a single data source, there can be interactions among different features. For instance, in the quantitative data, the price movements of two stocks and their corresponding industries can be highly correlated. It is obvious that the stocks affiliated to the same industry tend to co-evolve more frequently than those from unrelated industries. Although a tensor-based computational framework has been developed to model the joint impacts of different information sources for stock prediction~\cite{li2015tensor,li2016tensor}, each stock prediction in that framework is modeled as an individual task and thus learned independently, regardless of the correlations among the stocks.




In this paper, we for the first time present a novel tensor-based computational framework that can effectively predict stock price movements by fusing various sources of information. To this end, we extensively collect the stock-related information that can roughly be categorized into three types, 
~the quantitative information (e.g., historical stock prices) from the financial data providers, the event-specific information from the Web media, and the sentiment information from the social media. Specifically, we first collect and process financial Web news to extract the stock-related events, and merge the similar events into one category. The merged events with smaller number of event types can alleviate the data sparsity issue to some extent. We then process stock-related posts from a popular Chinese financial discussion board named Guba~\footnote{http://www.guba.com.cn}, and extract the sentiment indicators from users' posts and discussions with the sentiment classification technique. We next propose a coupled matrix and tensor factorization scheme to integrate the quantitative stock price data, the sentiment-specific data as well as the event-specific data. With the collaboratively factorized low rank matrices, we can effectively predict the stock movements by completing the missing values in the sparse tensor. The main innovation is that, in contrast to previous tensor-based studies which treat each stock prediction task independently, this scheme considers the correlations among the stocks and provides a powerful tool to co-learn all tasks simultaneously through the implicit shared knowledge and the explicit side information, and thus achieves better predictive performance.


The main contributions of this work can be summarized as follows:

\begin{itemize}
\item[1)]  We propose a stock price movement prediction framework by integrating multiple sources of information including Web news and social media postings. Compared with traditional methods, this framework considers the joint impacts of events and the public opinions on investment decisions.

\item[2)] To alleviate the data sparsity issue and utilize the commonalities among stocks, we explore the correlations among stocks with various methods. Instead of a simple linear combination of the stock-related features, we consider the coupling effects among features to capture their correlations.

\item[3)] We propose a coupled matrix and tensor factorization scheme to support the heterogeneous information integration and multi-task learning simultaneously. The stock price movement prediction is then completed by multiplying the factorized low rank factor matrices.

\item[4)] Our model is evaluated on two datasets, the China A-share stock market data and the HK stock market data, and the results show that our proposal can achieve the accuracy of 62.5\% and 61.7\% respectively. Compared to the state-of-art baseline, our method not only demonstrates superiority in performance, but also requires fewer parameters to tune. 
\end{itemize}


The remainder of this paper is organized as follows. Section 2 introduces the related work. We give the preliminary in Section 3. Section 4 describes the system framework. We elaborate the coupled matrix and tensor factorization scheme in Section 5. In Section 6, evaluation on real data shows the effectiveness of the proposed approach. Finally, we conclude the paper in Section 7.

\section{Related work}


Various studies have found that financial news can dramatically affect the stock price~\cite{cutler1988moves,tetlock2008more,luss2015predicting,xie2013semantic,wang2014semiparametric,peng2015leverage}. According to~\cite{kumar2016survey}, it summarizes several text mining methods that are used in the financial field. Structured events have been extracted as tuples including the agent, predicate and object from news documents in~\cite{ding2014using}. Deep learning-based method is proposed to learn event representations to capture syntactic and semantic information based on word embeddings~\cite{ding2015deep}. Recently, the external information from knowledge graphs is incorporated into the learning process to generate event representations~\cite{dingknowledge}. A deep neural model is proposed to measure the information content of financial news, aiming to understand an event's economic value~\cite{changmeasuring}. Temporal properties of news events that have short- or long-term influences on stock prices are modeled in~\cite{yoshihara2014predicting}. 
However, these works only consider the impacts of news events but ignore the fact that the moods of investors may also lead to stock price fluctuations.


There are also a line of researches trying to apply sentiment analysis on information sources to analyze the impacts of sentiments on stock market volatility. One main data source is the news articles~\cite{nguyen2015topic,feldman2011stock,schumaker2009quantitative,schumaker2009textual,li2014news}. A hybrid approach is introduced in~\cite{feldman2011stock} for stock sentiment analysis based on news articles on companies. In~\cite{wang2012novel}, textual data are represented as feature vectors by using the bag-of-words approach, and the data are organized as the input of a time series model, showing the use of market sentiments can improve the forecasting accuracy. The role of author tone in financial news articles is investigated in~\cite{schumaker2012evaluating}, which suggests a contrarian investment manner, i.e., see good news, sell; see bad news, buy. 
~The time series data of the Web information in a day is built by summing the sentiment values of the words in all the articles, and a support vector regression (SVR) is used to build the mapping from the Web information time series to the price values~\cite{liang2013associating}. News sentiments are measured and the combined effect of Web news and social media on stock markets are studied in~\cite{li2014media} .

Another widely used data source to extract sentiments is the social media~\cite{nguyen2015topic,zhang2011predicting,si2013exploiting}. It has been proved that the public mood extracted from Twitter can have impacts on stock market volatility, where mood tracking tools, such as OpinionFinder and Google Profile of Mood States, are used to analyze the text content of daily Twitter~\cite{bollen2011twitter}. A method to measure the collective hope and fear on each day and analyze the correlation between these indices and the stock market indicators is proposed in~\cite{zhang2011predicting}, by which each tweet is tagged with mood words as fear, sorry, hope and so on. The results show that the ratio of the emotional tweets~present a significant negative correlation with Dow Jones, NASDAQ and S\&P 500. A topic-based sentiment time series approach is proposed to predict the market~\cite{si2013exploiting}. This work is extended to further exploit the social relations between stocks from the social media context. A stock network is built with Twitter~based on the co-occurring relationships, and a labeled topic model is employed to jointly model the tweets and the network structure to assign each node and each edge a topic respectively. Then, a lexicon-based sentiment analysis method is used to compute the sentiment score for each node and each edge topic. Finally, the sentiment time series and price time series are used for prediction~\cite{si2014exploiting}. Though these works have shown their effectiveness by incorporating sentiments, they ignore other important factors such as news events in their model.

In addition to the aforementioned studies that consider either the events or the sentiments separately, recent studies begin to investigate their joint impacts. How news articles affect the trading activities is studied in~\cite{li2014effect}, and the sentiment impacts are also testified. The closest approach to ours resides in~~\cite{li2015tensor,li2016tensor}, where events and sentiments are also integrated in a tensor framework together with firm-specific features (e.g., P/B, P/E), to model the joint impacts on the stock market volatility. However, they build a tensor for each stock in each day, which leads to a set of drawbacks. Firstly, with that method, each stock requires a set of parameters to set, and thus it is a parameter-laden approach due to the large volume of stocks in market, making it difficult to achieve robustness across various scenarios without careful tuning. Secondly, they don't take into account the relationships that exist among the stocks, and thus cannot utilize their commonalities to help each task be learned better. Last, they don't address the data sparsity issue. By contrast, the problem deteriorates as they build a time series tensor stream, i.e., a tensor is built for each stock in each time unit (e.g., day or hour).
 Apart from events and sentiments, there are also some works extracting other features from Web information. For example, the relationship between Twitter activities and stock market is studied under a graph-based view~\cite{ruiz2012correlating}. Specifically, each collection of tweets from a company is represented as a graph, from which activity-based (e.g., the number of hashtags) and graph-based features (e.g., the link-structure) are extracted for cross correlation coefficient analysis. However, the results show that the prices of a stock are weakly correlated with the analyzed features. By combining analysts' recommendations on the Web and stock returns, a posterior probability model is built based on Bayesian inference~\cite{DUAN2013151}. 
 
There are also a few studies in the field of behavioral finance that discuss the effects of overconfidence in the decision-making process of investors~\cite{behaviorbook,maplosone}, but the impacts of social media on the overconfidence are still under exploration. Previous research on psychological-overconfidence theory indicates that investors tend to overweight private information while ignoring public information~\cite{Chuang2006An}. Hence, the overconfidence may be alleviated as the investors are well informed with social media. The other viewpoint is that social media may strengthen the overconfidence. As the diverse contagions in social media, investors may just seek the opinions that are consistent with their prior beliefs and thus confirm their beliefs, while they tend to consciously ignore adverse opinions~\cite{behaviorbook}. Thus, the relationship between the social media and overconfidence is very complicated and how to measure the overconfidence quantitatively is related to psychology and beyond the scope of this paper, but it is a promising research direction and can be our future work.

Stock correlations are important to better understand the stock market, and can be measured in various ways. A model of coupled random walks is proposed in~\cite{PhysRevE.70.026101} to model the stock correlations, and the walks are coupled via the mechanism that the price change of each stock is activated by the price gradients over some underlying network. The correlations between stocks are reflected by the correlations between Wiener process and a stochastic correlation model is proposed in~\cite{Chen938934}. Stock correlations as a time series are modeled as a mean reverting process, together with a term related to index return~\cite{sepp2011modeling}. In~\cite{preis2012quantifying}, the average correlations among stocks are found to scale linearly with the market stress reflected by normalized DJIA index returns on various time scales. However, these existing works use different stock correlation measures from our proposal. Specifically, they don't apply coupled stock similarity measure, nor use the user perceived correlations extracted from social network as proposed in this paper. Moreover, different from our proposal, the correlations they obtain didn't be used in the coupled matrix and tensor framework to address the sparsity issue in tensor decomposition. Please also note that our framework is general and stock correlations obtained with other methods can also be fed into this framework.

To summarize, different from most of existing studies that only consider either news events or sentiments in their models, our proposal involves both of these two crucial driving factors, and effectively integrates them together with historical quantitative data into a novel coupled matrix and tensor framework. In addition, we explore the stock correlations to facilitate the tensor decomposition, aiming to achieve better stock predictive power, which is rarely covered by previous studies.

%
%
%
%
%
%
%
%
%
%

\section{Preliminaries}

\subsection{Tensor decomposition and reconstruction}

In this part, we briefly introduce the mathematical notations and the tensor operations used in this paper. Tensors are high-order arrays that generalize the notions of vectors and matrices. In this paper, we use a third-order tensor, representing a 3-dimensional array. Scalars are 0th-order tensors and denoted by lowercase letters, e.g., $a$. Vectors are 1st-order tensor and denoted by boldface lowercase letters, e.g., $\textbf{a}$. Matrices are 2nd-order tensor and denoted by boldface capital letters, e.g., $\bm{X}$, and 3rd-order tensors are denoted by calligraphic letters, e.g., $\mathcal{A}$. The $i$th entry of a vector $\bm{a}$ is denoted by $a_i$, element $(i, j)$ of a matrix $\bm{X}$ is denoted by $x_{ij}$, and element $(i,j,k)$ of a 3rd-order tensor $\mathcal{A}$ is denoted by $a_{ijk}$. The $i$th row and the $j$th column of a matrix $\bm{X}$ are denoted by $\bm{x}_{i:}$ and $\bm{x}_{:j}$, respectively. Alternatively, the $i$th row of a matrix, $\textbf{a}_{i:}$, can also be denoted as $\textbf{a}_i$ .

The norm of a tensor $\mathcal{A} \in \mathcal{R}^{N \times M \times L}$ is defined as:
\begin{equation*}
\centering
\begin{split}
\lVert{\mathcal{A}}\rVert = \sqrt{\sum_{i=1}^N \sum_{j=1}^M\sum_{k=1}^L a_{i,j,k}^2}
\label{tensor1}
\end{split}
\end{equation*}
This is analogous to the matrix \emph{Frobenius norm}, which is denoted as $\lVert{\bm{X}}\rVert$ for a matrix $\bm{X}$. The \emph{n-mode product} of a tensor $\mathcal{C} \in \mathcal{R}^{I_1 \times I_2 \times I_3}$ with a matrix $\bm{U} \in \bm{R}^{I_n \times J}$, denoted by $\mathcal{C} \times_n \bm{U}$, is a tensor of size ${I_1 \times ... \times I_{n-1} \times J \times I_{n+1} \times ... \times I_N}$ with the elements $(\mathcal{C} \times_n \bm{U})_{i_1...i_{n-1}j\i_{n+1}...i_N}=\sum_{i_n=1}^{I_n}{a_{i_1,i_2,i_N}u_{i_nj}}$.

The \emph{Tucker factorization} of a tensor $\mathcal{A} \in \mathcal{R}^{N \times M \times L}$ is defined as:
\begin{equation*}
\centering
\begin{split}
\mathcal{A}= \mathcal{C} \times_1 \bm{U} \times_2 \bm{V} \times_3 \bm{W}
\label{tensor2}
\end{split}
\end{equation*}
Here, $\bm{U} \in \mathcal{R}^{N \times R}$, $\bm{V} \in \mathcal{R}^{M \times S}$ and $\bm{W} \in \mathcal{R}^{L \times T}$ are the factor matrices and can be thought of as the principal components in each mode. The tensor $\mathcal{C} \in \mathcal{R}^{R \times S \times T}$ is the core tensor and its entries show the level of interaction between the different components. The reconstructed tensor is derived by multiplying the core tensor and the three factor matrices. It can be observed that tensor decomposition and reconstruction has updated the value for each existing entries indicating its importance and fill some new entries showing the latent relationships. Generally speaking, tensor factorization can be regarded as an extension of the matrix decomposition. During the process of decomposition, data can be projected in the subspaces, which includes latent significance.

\subsection{Coupled attribute value similarity}
\label{sec:cos}

This paper exploits the stock correlations to facilitate the prediction, and similarity measures are designed to obtain such correlations. The traditional similarity measures usually assume the object's attributes are independent with each other and the interactions among the attributes are not considered to calculate the similarity. However, the coupling effects among different attributes exist in a wide range of applications. We take two movie-related attributes, the actor and genre, as an example. The actor Jackie Chan appears more frequently in action movies than in other genres, while actor Jim Carrey is more likely to play roles in comedy movies. Thus, in addition to considering the Intra-coupled similarity within an attribute, the Inter-coupled similarity among different attributes should also be involved.
%
%

Formally, a large number of data objects with the same features can be organized by such information table $S=<U,A,V, f>$ , where $U=\{u_1,u_2,...,u_m\}$ is a set of instances, $A=\{a_1,a_2,...,a_n\}$ is a $n$-attribute set for every instance. $V_j$ is the set of all the values of feature $a_j$, and $f_j:U \to V_j$ is a map function that returns the attribute $a_j$'s value of the instance. We next introduce an approach to calculate the Intra-coupled and Inter-coupled similarities.


\textbf{Intra-coupled attribute value similarity (IaAVS).} According to~\cite{wang2011coupled}, the frequency distribution of the attribute value can reveal the value similarity. The Intra-coupled similarity $\delta_j^{I_a}(x,y)$ between values $x$ and $y$ of attribute $a_j$ is defined as:

\begin{equation*}
\centering
\begin{split}
\delta_j^{I_a}=\frac{\rvert g_j(x) \rvert \cdot \rvert g_j(x) \rvert}{\rvert g_j(x) \rvert+\rvert g_j(y) \rvert+\rvert g_j(x) \rvert \cdot \rvert g_j(y) \rvert}
\label{}
\end{split}
\end{equation*}

\noindent where $g_j:V_j \to 2^U$ is a map function that returns a set of instances whose values of attribute $a_j$ are $x$. Thus, $g_j(x)$ is defined as:

\begin{equation*}
\centering
\begin{split}
g_j(x)=\{u_i \rvert f_i(u_i)=x,1 \le j \le n,1 \le i \le m\}
\label{}
\end{split}
\end{equation*}

\textbf{Inter-coupled attribute value similarity (IeAVS).} The Inter-coupled attribute value similarity $\delta_j^{I_e}(x,y)$ between values $x$ and $y$ of attribute $a_j$ is the aggregation of the relative similarities $\delta_{j \rvert k}(x,y)$ (which will be given later) for all the other attributes excluding itself.

\begin{equation*}
\centering
\begin{split}
\delta_j^{I_e}(x,y)=\sum_{k=1,k \ne j}^n \alpha_k \delta_{j \rvert k}(x,y)
\label{}
\end{split}
\end{equation*}
where $\alpha_k$ is the weight parameter for attribute $a_j$, $\sum_{k=1}^n \alpha_k=1$, $\alpha_k \in [0,1]$. The relative similarity $\delta_{j \rvert k}(x,y)$ represents the similarity between values $x$ and $y$ of attribute $a_j$, based on the other attribute $a_k$. Thus we have

\begin{equation*}
\centering
\begin{split}
\delta_{j \rvert k}(x,y)=\sum_{w \in \bigcap} min\{P_{k \rvert j}(\{w\}\rvert x),P_{k \rvert j}(\{w\}\rvert y)\}
\label{}
\end{split}
\end{equation*}

\noindent where  $w \in W$, and $W$ is the $k$th attribute value subset and thus $W \subseteq V_k$. $w\in\bigcap$ denotes $w\in\varphi_{j\rightarrow{k}}(x)\bigcap\varphi_{j\rightarrow{k}}(y)$, and $\varphi_{j\rightarrow{k}}\colon{V_j}\rightarrow{2^{V_k}}$ is a map function that returns the attribute $a_k$'s values subset of the instances, whose attribute $a_j$'s values are $x$, and that is,

\begin{equation*}
\centering
\begin{split}
\varphi_{j\rightarrow{k}}(x)=f_k^*(g_j(x))
\label{}
\end{split}
\end{equation*}

\noindent where $f^*(\cdot)$ differs from $f(\cdot)$ in that the input of the function $f^*(\cdot)$ is a set of instances instead of an individual instance. $P_{k \rvert j}(\{w\} \rvert x)$ is the Information Condition Probability of $\{w\}$ with respect to $x$. $P_{k \rvert j}(W \rvert x)$ can be obtained through

%

\begin{equation*}
\centering
\begin{split}
P_{k \rvert j}(W \rvert x)=\frac{\rvert g_k^*(W) \bigcap g_i(x) \rvert}{\rvert g_i(x) \rvert}
\label{}
\end{split}
\end{equation*}

\noindent Here, $g_k^{*}(W)$ is the variation of $g_k(x)$ with the set $W$ as input. Specially, $g_k^{*}(W)$ maps a set of attribute $a_k$ values $W$ to a set of instances, that is 

\begin{equation*}
\centering
\begin{split}
g_k^{*}(W)=\{u_i \rvert f_i(u_i)\in{W},1 \le k \le n,1 \le i \le m\}
\label{}
\end{split}
\end{equation*}

For more detailed introduction to the coupled attribute value similarity, one can refer to~\cite{wang2011coupled,Li2015}.

\section{The system framework}
In this section, we depict the proposed system framework for stock price movement prediction. Specifically, we first introduce how to extract the effective features from multiple data sources and then show how to combine these features into the prediction model.

\begin{figure}
\setlength{\abovecaptionskip}{0.02pt} 
\centering
\epsfig{file=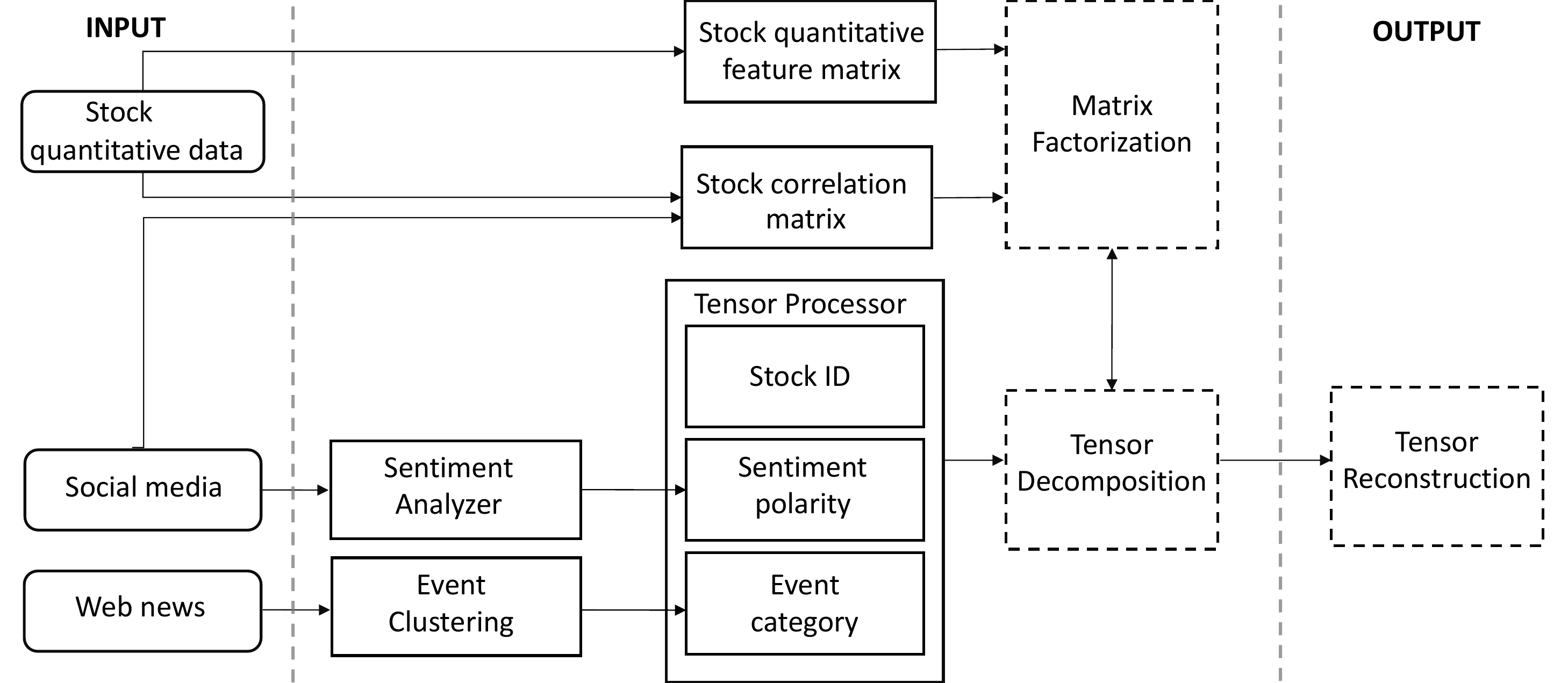, height=2.4in, width=4.8in}
\caption{The system framework of our stock prediction model}
\label{Framework}
\end{figure}

We employ the historical stock quantitative data, Web news articles and social media data to construct a third-order tensor as well as two auxiliary matrices to model their joint effects on stock price movements. The overall system framework is shown in Fig.~\ref{Framework}, which is comprised of four major parts: 
~1) The stock quantitative feature matrix construction based on the stock quantitative data;  2) The stock correlation matrix construction with the multi-sourced data; 3) Events and sentiments extraction from news articles and social media to build the stock movement tensor; 4) Coupled matrix and tensor factorization for stock price movement prediction. The reconstructed tensor would be the output of the system and used to make stock predictions. We will describe the first three parts in this section, and the last part will be introduced in the next section.

\subsection{Building the stock quantitative feature matrix}

The first step is to build the stock quantitative feature matrix whose two dimensions are the stock IDs and the quantitative features respectively. For every stock $i$, the quantitative features are denoted as such a vector $x_i=(x_{i1},x_{i2},...,x_{ik},...,x_{iK})$, where $K$ is the number of features, $x_{ik}$ is the value of the $k$th feature. Then we normalize and gather all the features to form the stock quantitative matrix $\bm{X} \in \mathcal{R}^{N \times K}$ , where $N$ is the number of stocks.

Based on the previous studies~\cite{fama1993common,li2015tensor}, we choose four common quantitative features, that is, the share turnover, the price-to-earning (P/E) ratio, the price-to-book (P/B) ratio and the price-to-cash-flow (PCF) ratio. Share turnover is a measure of stock liquidity calculated by dividing the total number of shares traded over a period by the average number of shares outstanding for the period. The higher the share turnover, the more liquid the share of the company. Our investigation on the social media shows that stocks with a larger number of discussions usually attract more investors' attention and have a higher turnover. P/E ratio is quite informative in the stock market, indicating the dollar amount an investor can expect to invest in a company in order to receive one dollar of that company's earnings. In general, a higher P/E suggests that investors are expecting higher earnings growth in the future. P/E ratio thus can be considered as a benchmark and used to identify whether or not a stock is worth buying. Note that during a period, the average P/E ratio of a firm is relatively stable, with fluctuations depending on economic conditions, but P/E may vary widely for different industries. Thus, similar companies usually exhibit similar P/E ratios. P/B ratio is similar to P/E ratio which reflects the intrinsic value of a company, but is used when P/E is distorted in some situations, e.g., for financial industry where the market value is heavily related to the value of its equity. PCF is also a common indicator of a stock's valuation, especially useful for valuing stocks that have positive cash flow but are not profitable. 

Each of the above quantitative features can reflect the status and valuation of a company to some extent from an aspect, and the combination of them could depict a company's valuation from a comprehensive perspective. That's why we construct the matrix with various quantitative features. After factorizing this matrix, each stock will be represented by a vector (embedding), and it can be expected that similar stocks will have close quantitative feature vectors. 

\subsection{Building the stock correlation matrix}

In the stock market, stocks are usually not independent of each other, and they can be correlated from various perspectives. For example, they may belong to the same industry, or they are involved in the same topic (e.g., benefit from reducing interest rate), or just co-evolves in price historically without explicit relationships. The correlations between two stocks can be depicted by their similarities. Traditional similarity metrics usually require that the objects are described by numerical features, and measure their similarity by the geometric analogies which reflect the relationship of data values. However, the underlying assumption of these metrics is that the features follow the independent and identically distribution (iid), indicating they only consider the intra-coupled similarity within a feature, but ignore the dependency relationships among features~\cite{wang2011coupled}. In this paper, we extract the coupled correlations between stocks by considering the coupled effects among features, which is denoted as the coupled stock correlation. For the purpose of comparison, we also develop three other methods to calculate the correlations, which are described in detail as follows.

%


\textbf{Coupled stock correlation.} In complex applications such as stock analysis, there usually exists coupling effects between different features. To capture such effects, inspired by the previous works~\cite{wang2011coupled,Li2015}, we apply a Coupled Stock Similarity (CSS) measure by taking the intra interaction between values within an attribute and the inter interaction between attributes into account to calculate the correlation between stocks. In this work, the coupling attributes of stocks include the closing price and the industry index trend in each trading day, which are both crucial attributes for a stock, and empirically have coupling effects. In particular, a stock's price usually exhibits similar fluctuation trends as its industry index. 

\begin{figure}
\centering
\epsfig{file=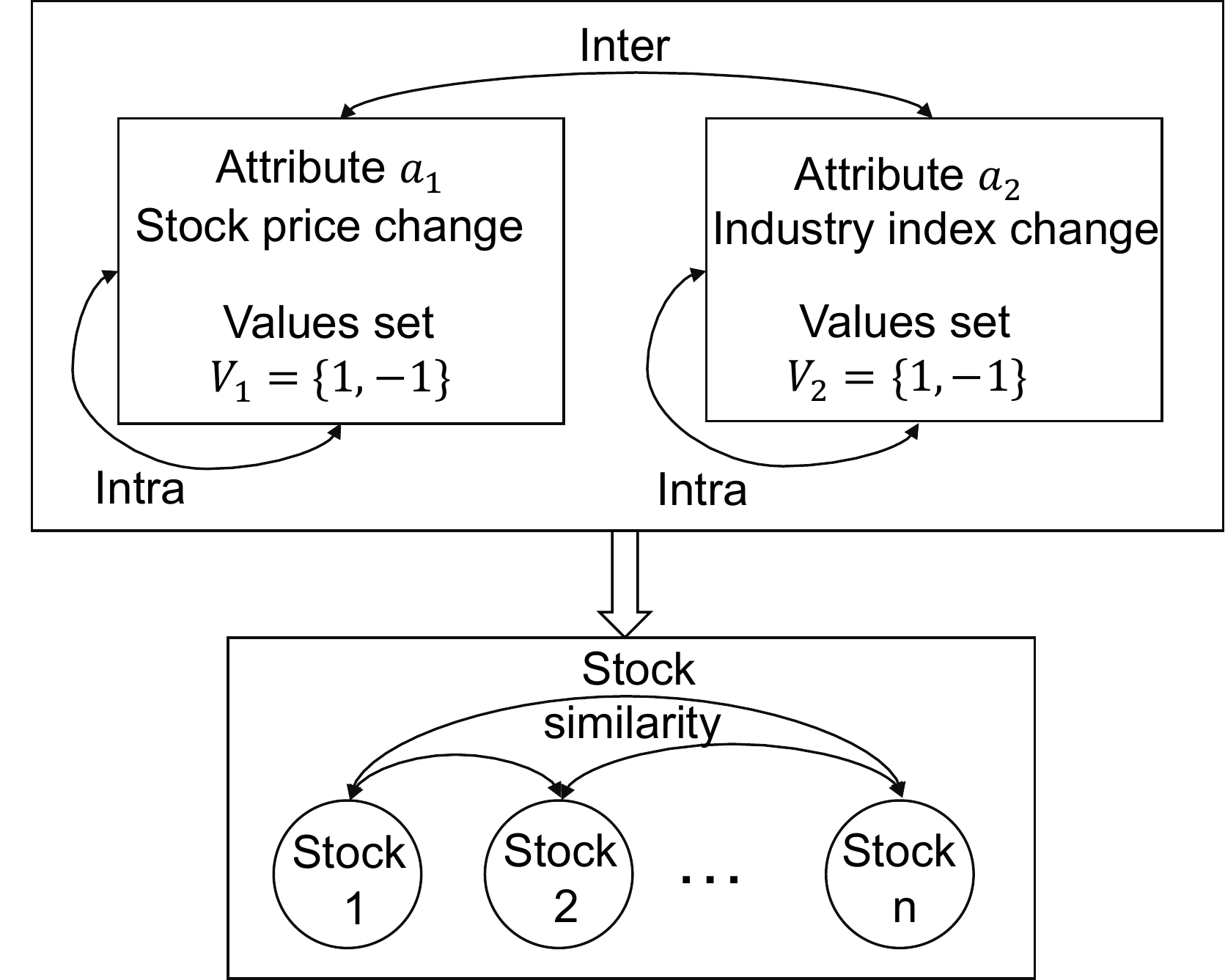, height=2.5in, width=3in}
\caption{Coupled stock similarity}
\label{Feature}
\end{figure}

The idea of the Coupled Stock Correlation is depicted in Fig.~\ref{Feature}. Given the stock attribute space $S_a =\left \langle S, A, V, f \right \rangle$, where $S=\{s_1,s_2,...,s_n\}$ is a set of stocks, $A=\{a_1,a_2,...,a_m\}$ is the attribute set of the stock, $V_k$ is the set of all the values of the feature $a_k$, $V_{ik}$ is the value of feature $a_k$ for stock $s_i$, and $f_k:S \to V_k$ is a map function that returns the attribute $a_k$'s value of the stock. Then CSS between two stocks $s_i$ and $s_j$ can be defined as

\begin{equation}
CSS(i,j) = \sum_k\delta_{k}^{I_a}(V_{ik},V_{jk})*\delta_{k}^{I_e}(V_{ik},V_{jk})
\label{eqn:css}
\end{equation}

\noindent where $V_{ik}$ and $V_{jk}$ are the values of feature $k$ on stock $s_i$ and $s_j$ respectively, $\delta_{k}^{I_a}(V_{ik},V_{jk})$ is the intra-coupled attribute values similarity of attribute $a_k$,  and $\delta_{k}^{I_e}(V_{ik},V_{jk})$ is the inter-coupled attribute values similarity which can be calculated based on other coupled attributes. The details of the theoretical analysis and calculation method of Eq. (1) can be referred to Sec.~\ref{sec:cos} and literature~\cite{wang2011coupled}.

To derive $CSS(i,j)$, the attribute set used can be represented as a tuple $s_{it}=(p_{it},c_{it})$, where  $s_{it}$ is the status of stock $i$ on day $t$.  $s_{it}$ has two attributes: (1) $p_{it}$ the price movement direction of the stock $i$ on day $t$, and 1 means going up while -1 means going down; (2) $c_{it}$ represents the industry (that sock $i$ belongs to) index changing direction on day $t$. Analogously, the value of $c_{it}$ is 1 for going up and -1 for going down. Then we calculate $CSS(i,j)$ for each pair of stocks $s_i,s_j$ on each day, and average all the $CSS(i,j)$ through the training period (nine months in our case).

\textbf{Co-evolving direction correlation.} We also attempt to learn the correlation between two stocks by simply counting the times of co-evolvements in their price movements. In particular, for each pair of stocks, if their closing prices both go up (or go down) compared to their closing prices in the previous trading day, they are referred as co-evolving at that day. Given a time period (e.g., a year), the more days the two stocks co-evolves, the more closely they are correlated. Formally, supposing the number of days on which the two stocks $s_i$ and $s_j$ co-evolve is $N$, and the total number of trading days is $M$, the correlation coefficient between them is $N/M$, which is also the value of entry $z_{ij}$ in the stock correlation matrix $\bm{Z}$.

\textbf{Co-evolving p-change correlation.} The p-change value of a stock on the trading day $i$ is defined as the change rate between its closing prices on the day $i$ and on the previous day $(i-1)$. By combining each stock's p-change for all the trading days in a time period, we can obtain a p-change curve for each stock. Then Pearson Correlation Coefficient is applied to measure the co-evolving p-change correlation of each pair of stocks based on their p-change curves. This measurement considers both the fluctuation direction and the fluctuation range to reflect the co-evolving movements.
 
\textbf{User perceived correlation.} In addition to obtaining the correlations with the stock quantitative data, we can also extract the user perceived relatedness through Xueqiu, which is a Twitter-like investor social network in China~\footnote{https://xueqiu.com/}. To obtain such correlations, we collect all the pairwise stocks mentioned by the same tweet and then remove the tweets mentioning more than five consecutive stock tickers as such tweets usually do not convey useful meaning for our task.

With the above methods, the correlation coefficients for each pair of stocks are obtained. Next they are normalized and filled into the stock correlation matrix $\bm{Z} \in \mathcal{R}^{N \times N}$, whose entry $z_{ij}$ is the normalized correlation value between stock $i$ and stock $j$. A larger $z_{i,j}$ means a higher correlation between the stocks $i$ and $j$. Please note that each individual method described above can provide a correlation matrix as the input of our coupled matrix and tensor factorization model, and we will empirically choose the best one according to the evaluating results in the experimental section.

\subsection{Building the stock movement tensor}
\label{sec:buildtensor}

We use a tensor to represent the stock ID, events and sentiments collected from multiple data sources. The reason is that the consequence of an event is usually complex and thus only relying on events is not sufficient to make good predictions. For example, the announcement of acquisition can either be positive or negative news under different circumstances. To address this issue, the sentiments extracted from social media, which represent the people's judgments on the events, can be an effective complement. Specifically, the positive sentiment usually indicates the event is good to the stock while the negative sentiments mean the opposite.  

Formally, we build a third-order tensor $\mathcal{A} \in \mathcal{R}^{N \times M \times L}$, whose three dimensions are stock ID, event category and social sentiment polarity. Note that no events happening is also regarded as a type of event. We next show how to extract the events and sentiments from Web news and social media respectively.

\textbf{Event extraction.} Previous work~\cite{ding2014using} show that news titles are more useful to extract events compared to news contents. Thus, this paper extracts events only from news titles. Within a title, we use the verb or gerund to represent an event since they are quite informative. For example, in the news title ``Microsoft to acquire LinkedIn", the verb "acquire" can denote the event quite well. Note that we do not consider the subject or object in the title as in our Web news data source, the news has been assigned to the related stocks, which will be the subject or object of the event. The data source will be described in Sec.~\ref{datacollection}.

To extract the events, we first segment the news titles into words with Jieba~\footnote{https://github.com/fxsjy/jieba}, an open-source python component for Chinese text segmentation. For each part of the speech tagging, we next extract the verbs and gerunds in news titles. If we directly use the extracted verbs and gerunds (more than 6,000 in our case) to construct the tensor, the tensor will be extremely sparse, preventing it from getting a good prediction accuracy. We observed that a lot of titles actually refer to the same type of event but fall into different event categories. For example, two titles ``Microsoft to acquire LinkedIn" and ``Microsoft to buy LinkedIn" can be treated as the same event, but are presented by different verbs. To address this issue, we examine the corpus of the extracted verbs and gerunds, and cluster the synonyms with a linguistic knowledge base HowNet~\cite{hownet}. To further reduce the dimensionality of the event categories, we next cluster the event categories based on their word embeddings. Specifically, we train the domain-specific word embeddings using word2vector~\cite{mikolov2013distributed} with the Chinese finance news corpus~\cite{zhang2016effective}, and the number of dimensions is set as 100. We then apply the $k$-means method to cluster the embeddings and obtain 500 clusters, which is set empirically. Too many clusters may lead to over sparsity of the tensor, while too few clusters may not be sufficient to separate different types of events into different categories. Please note that the word embedding can't distinguish antonyms, i.e., the antonyms may fall into the same cluster. For example, in the context of stock market, the embedding vectors of ``rise'' and ``fall'' may be close to each other (and thus in the same cluster), but they indicate totally different meanings for stock movements. Thus, manual correction is required to separate the antonym words within the same cluster into different clusters. As there are 97 antonym words in the same clusters, we create 97 new clusters and eventually the news titles are classified into 597 event categories.

%
%
%

\textbf{Sentiment extraction.} For every stock, we analyze its public sentiment polarity (i.e., positive or negative) on each day by extracting users' postings from the investor social media. We apply the method proposed in~\cite{li2015tensor} to calculate the public sentiment, and this method mainly utilizes the following information: each posting's publication time, the title, the number of clicks and the number of comments. Different from~\cite{li2015tensor}, we develop a specialized sentiment dictionary focusing on financial social media based on NTUSD~\cite{ku2007mining}. The new sentiment dictionary contains plenty of words with the sentiment polarity in the domain of finance, e.g., rise, fall, up, and down. To obtain the sentiment polarity of a posting, we first use Jieba to segment the postings, and extract the sentiment words by using the sentiment dictionary. We then calculate the positive and negative sentiment value for each stock on each day. The sentiment value is calculated by

\begin{equation*}
S_{it}^+ = \sum_{j=0}^K\frac{P_{jt}}{L_{jt}}\times W_{jt}
\label{eqn:topic-inter}
\end{equation*}

\noindent where $S_{it}^+$ is the positive sentiment value of stock $i$ on day $t$, $P_{jt}$ is the number of positive sentiment words in posting $j$ published on day $t$ for stock $i$, $L_{jt}$ is the total number of sentiment words in posting $j$ on day $t$, and $W_{jt}$ is the weight of posting $j$, which indicates the degree of impact on social media and can be calculated by the number of clicks and comments. The detailed calculation method can be referred to~\cite{li2015tensor}. Finally, the sentiment polarity of a stock on a day can be obtained by comparing the difference between its positive and negative values with a predefined threshold.


After extracting the events and sentiments for each stock on each day, we can construct a stock movement tensor for all the stocks on each trading day. A positive (negative) value of the entry ($a_{nml} = 1~or~-1$) of the tensor denotes the price of stock $n$ goes up (down) when event $m$ happens and simultaneously the public sentiment is $l$. However, due to the sparsity of the events for one stock, the tensor is overly sparse. Thus, we aggregate the tensors in a long past period to form a denser historical tensor in a period into a historical tensor. Specifically, the corresponding entry values in each day's tensor will be aggregated to form an upward probability, indicating the probability that a stock price will go up when the stock meets a specific event category and a specific sentiment polarity. For example, given ten tensors in the past ten trading days, if the entry $a_{nml}$ has six ``+1" and four ``-1" in all the ten tensors, we say its upward probability is 0.6. After aggregation, as the stock movement tensor is still sparse for an accurate decomposition, we then apply the stock correlation matrix and stock quantitative feature matrix to assist its decomposition, which will be described in the next section.

\section{Coupled matrix and tensor factorization}

%

In the above section, we have shown how to build the stock movement tensor. Although several techniques have been applied to reduce the dimension of the events category, the tensor is still very sparse, as the number of events related to each stock is quite limited. Consequently, solely decomposing the tensor does not work very well on making very accurate predictions. To address this issue, auxiliary information from other data sources can be incorporated to assist. In this work, the additional information includes the stock correlations and the stock quantitative features, which reside in two matrices, the stock quantitative feature matrix \bm{$X$} and the stock correlation matrix \bm{$Z$}. The main idea of this coupled model is to propagate the information among \bm{$X$}, \bm{$Y$}, and \bm{$Z$} by requiring them to share the low-rank matrices in a collective matrix and tensor factorization model. We can also illustrate this model from a multi-task learning perspective, that is, instead of conducting each stock prediction task independently, we can co-learn multiple tasks simultaneously through their commonalities and shared knowledge. In our work, the multiple tasks are connected by the stock correlations and their quantitative features. The intuition behind is that if two stocks are highly correlated, the events occurred on one stock are likely to have similar effects on the other stock. 
 
%
%

%

\begin{figure}
\centering
\epsfig{file=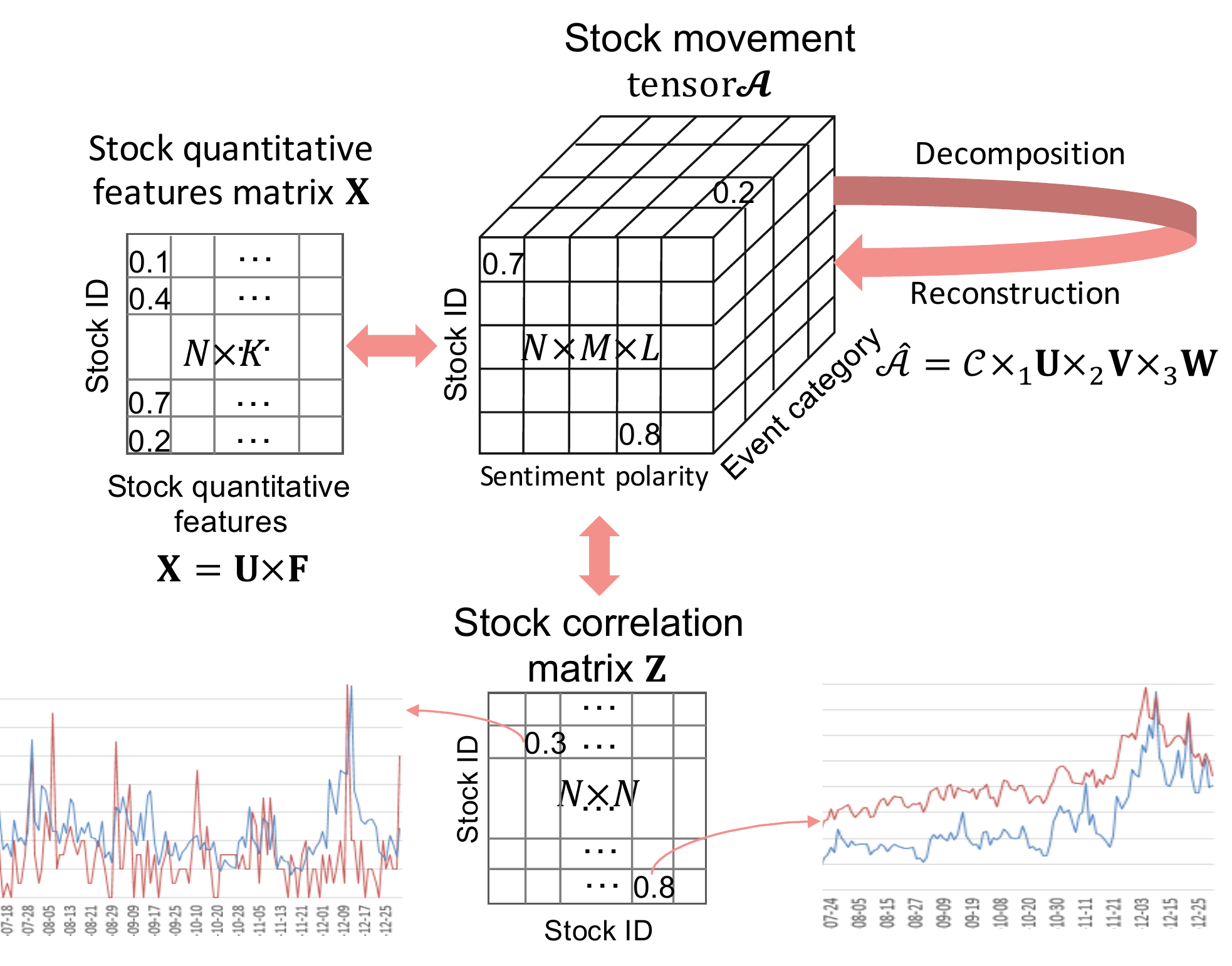, height=3.4in, width=4.4in}
\caption{Coupled matrix and tensor factorization}
\label{Model}
\end{figure}

We next describe how to collaboratively factorize matrices and tensors. Specifically, given a very sparse stock movement tensor $\mathcal{A}$, we try to complete $\mathcal{A}$ by decomposing it collaboratively with the stock quantitative feature matrix \bm{$X$} and the stock correlation matrix \bm{$Z$}. As shown in Fig.~\ref{Model}, $\bm{X} \in \mathcal{R}^{N \times K}$ is the stock quantitative feature matrix and $\bm{Z} \in \mathcal{R}^{N \times N}$ is the stock correlation matrix, where $N$ is the number of stocks, $M$ is the number of event types (categories), $L$ is the number of sentiment polarities, i.e., positive and negative, and $K$ is the number of quantitive features. The tensor $\mathcal{A}$ can be decomposed as $\mathcal{C} \times_1 U \times_2 V \times_3 W$, where the core tensor is $\mathcal{C} \in \mathcal{R}^{R_1 \times R_2 \times R_3}$ and three factorized low-rank matrices are $\bm{U} \in \mathcal{R}^{N \times R_1}$, $\bm{V} \in \mathcal{R}^{M \times R_2}$, $\bm{W} \in \mathcal{R}^{L \times R_3}$, denoting the low rank latent factors for stocks, events and sentiments, respectively. $\bm{X}$ can be factorized as  $\bm{X} = \bm{U} \times \bm{F}$, where $\bm{F} \in \mathcal{R}^{R_1 \times K}$ is the low rank factor matrix for quantitative features. 

As our model applies to the coupled matrix and tensor factorization to replenish the tensor, the entries obtained after reconstruction are required to be close to their real values. To achieve this goal, we define the following objective function to minimize the factorization errors.


\begin{equation}
\begin{split}
\mathcal{L}(U,V,W,\mathcal{C},F) = \frac{1}{2} \lVert{\mathcal{A}-\mathcal{C} \times_1 U \times_2 V \times_3 W}\rVert^2 
 + \frac{\lambda_1}{2} \lVert{X-UF}\rVert^2\\ 
 + \frac{\lambda_2}{2} {tr(U^TL_ZU)}
 + \frac{\lambda_3}{2} (\lVert{U}\rVert^2 + \lVert{V}\rVert^2 + \lVert{W}\rVert^2 + \lVert{\mathcal{C}}\rVert^2 + \lVert{F}\rVert^2)
\label{eqn:topic-inter}
\end{split}
\end{equation}

\noindent where $\lVert{\mathcal{A}-\mathcal{C} \times_1 U \times_2 V \times_3 W}\rVert^2$ is to control the decomposition error of the tensor $\mathcal{A}$, $\lVert{X-UF}\rVert^2$ is to control the factorization error of $\bm{X}$, $tr(\cdot)$ denotes the matrix traces, $\lVert{U}\rVert^2 + \lVert{V}\rVert^2 + \lVert{W}\rVert^2 + \lVert{\mathcal{C}}\rVert^2 + \lVert{F}\rVert^2$ is the regularization penalty to avoid overfitting. $\bm{L_z} = \bm{D}-\bm{Z}$ is the Laplacian matrix of the stock correlation graph in which $\bm{D}$ is a diagonal matrix with diagonal entries $d_{ii}= \sum_{i} z_{ij}$.  And ${tr(\bm{U}^T\bm{L_z}\bm{U})}$ can be obtained through Eq.~(\ref{eqtr}), in which two stocks $s_i$ and $s_j$ with a higher correlation (i.e., $z_{ij}$ is large) should also have a closer distance between the vector $\bm{u_i}$ and $\bm{u_j}$ in the matrix $\bm{U}$.

\begin{equation}
\begin{split}
&\frac{1}{2} \sum_{i,j}\lVert{u_i-u_j}\rVert_2^2 z_{ij} = \sum_{i,j}u_iz_{ij}u_i^T- \sum_{i,j}u_iz_{ij}u_j^T \\
&=\sum_iu_id_{ii}u_i^T-\sum_{i,j}u_iz_{ij}u_j^T \\
&=tr(\bm{U}^T(\bm{D}-\bm{Z})\bm{U})=tr(\bm{U}^T\bm{L}_z\bm{U})
\end{split}
\label{eqtr}
\end{equation}

The object function is not jointly convex to all the variables $\bm{U},\bm{V},\bm{W},\mathcal{C},\bm{F}$. Thus, we use an element-wise optimization algorithm to iteratively update each entry in the matrices and tensor independently by gradient descent~\cite{karatzoglou2010multiverse,wang2015citywide}. The gradient for each variable is derived as follows:

\begin{equation*}
\centering
\begin{split}
\nabla_{u_{i:}}\mathcal{L}&=(\mathcal{C} \times_1 u_{i:}^T \times_2 v_{j:} ^T\times_3 w_{k:}^T-a_{ijk})\mathcal{C} \times_2 v_{j:}^T \times_3 w_{k:}^T\\
&+\lambda_1(u_{i:}F-x_{i:})F^T+\lambda_2(L_ZU)_{i:}+\lambda_3u_{i:}\\
\nabla_{v_{j:}}\mathcal{L}& = (\mathcal{C} \times_1 u_{i:}^T \times_2 v_{j:}^T \times_3 w_{k:}^T-a_{ijk})\mathcal{C} \times_1 u_{i:}^T \times_3 w_{k:}^T+\lambda_3v_{j:}\\
\nabla_{w_{k:}}\mathcal{L}&=(\mathcal{C} \times_1 u_{i:}^T \times_2 v_{j:}^T \times_3 w_{k:}^T-a_{ijk})\mathcal{C} \times_1 u_{i:}^T \times_2 v_{j:}^T+\lambda_3w_{k:}\\
\nabla_{\mathcal{C}}\mathcal{L}&= (\mathcal{C} \times_1 u_{i:}^T \times_2 v_{j:}^T \times_3 w_{k:}^T-a_{ijk})u_{i:} \circ v_{j:} \circ w_{k:}+\lambda_3C\\
\nabla_{F}\mathcal{L}&= \lambda_1u_{i:}^T(u_{i:}F-x_{i:})+\lambda_3F
\label{}
\end{split}
\end{equation*}

The detailed algorithm of the learning process is shown in Algorithm 1.

\begin{algorithm}
\caption{Coupled Matrix and Tensor Factorization}
\textbf{Input:} Tensor $\mathcal{A}$, matrices $X$, $Z$, an error threshold $\varepsilon$\protect\\
\textbf{Output:} Low rank matrices $U$, $V$, $W$, $F$, core tensor $\mathcal{C}$
\begin{algorithmic}[1]
\State Set $\eta$ as step size of gradient descent, iteration time
\State Initialize $U \in \mathcal{R}^{N \times R_1}$, $V \in \mathcal{R}^{M \times R_2}$, $W \in \mathcal{R}^{L \times R_3}$, $F \in \mathcal{R}^{R_1 \times K}$, $Z \in \mathcal{R}^{N \times N}$ and core tensor $\mathcal{C} \in \mathcal{R}^{R_1 \times R_2 \times R_3}$ with small random values, $t = 0$
\State $d_{ii} = \Sigma_i z_{ij}$
\State \bm{$L_Z$} = $D - Z$
\For{each $a_{ijk} \neq 0$}
\State Get $\nabla_{u_{i:}}\mathcal{L}$, $\nabla_{v_{j:}}\mathcal{L}$, $\nabla_{w_{k:}}\mathcal{L}$, $\nabla_{\mathcal{C}}\mathcal{L}$, $\nabla_{\mathcal{F}}\mathcal{L}$
\State $u_{i:}^{t+1} = u_{i:}^{t} - \eta\nabla_{{u_{i:}^t}}\mathcal{L}$
\State $v_{j:}^{t+1} = v_{j:}^{t} - \eta\nabla_{{v_{j:}^t}}\mathcal{L}$
\State $w_{k:}^{t+1} = w_{k:}^{t} - \eta\nabla_{{w_{k:}^t}}\mathcal{L}$
\State $\mathcal{C}^{t+1} = \mathcal{C}^{t} - \eta\nabla_{\mathcal{C}^{t}}\mathcal{L}$
\State $\mathcal{F}^{t+1} = \mathcal{F}^{t} - \eta\nabla_{\mathcal{F}^{t}}\mathcal{L}$
\EndFor
\While{($Loss^t-Loss^{t-1} > \epsilon$)}
\State Do steps $5 - 12$ iteratively
\State $t = t + 1$
\EndWhile
\State \Return{$U$, $V$, $W$ and $F$}
\end{algorithmic}
\end{algorithm}

\section{Experiments}

%

\subsection{Data collection and description}
\label{datacollection}

We evaluate our proposed method on two datasets, the China A-share stock market data and HK stock market data, during Jan. 1, 2015 and Dec. 31, 2015. For the A-share market, we choose 78 stocks from the Chinese Stock Index (CSI) 100, and collect the corresponding event information and sentiment polarity from Web media and social media. The remaining 22 stocks are not included in the experiment due to the very limited information on Web during this period. For HK market, since there are much less retail investors in HK market than in A-share market, the number of tweets in social network related to HK stocks is not as many as those for A-share stocks. Thus, we only choose 13 hot stocks with relatively large number of tweets for experiments~\footnote{The codes of the 13 HK stocks are 0175, 0388, 0390, 0400, 0656, 0700, 1030, 1766, 1918, 2318, 2333, 3333 and 3968.}. Next we introduce how to collect the data in detail as follows:

\textbf{Quantitative data:} The quantitative data of stocks of the two datasets are both collected from Wind, a widely used financial information service provider in China. The indices we select are the stock turnovers, P/E ratios, P/B ratios and PCF ratios that are commonly used indices for stock trading and valuation, which comprise the stock quantitative feature matrix. In addition, we also collect the closing price and industry index to calculate the coupled stock correlation.

\textbf{Web news data:} We collect 76,445 and 7284 news articles for A-share stocks and HK stocks respectively from Wind, including the titles and the publication time in 2015. Each article is assigned to the corresponding stock. These Web news are originally aggregated by Wind from major financial news websites in China, such as \url{http://finance.sina.com.cn} and \url{http://www.hexun.com}. These titles are then processed to extract events as described in Sec.~\ref{sec:buildtensor}. The data is publicly available on~\cite{webnews}.

\textbf{Social media data:} The sentiments for A-share stocks are extracted from Guba. Guba is an active financial social media where investors can post their opinions and each stock has its own discussion site. In total, we collect 6,163,056 postings from Jan.1, 2015 to Dec. 31, 2015 for the 78 stocks. For each posting, the content, user ID, the title, the number of comments and clicks, and the publication time are extracted. We also have made this dataset publicly available~\cite{gubadata}. In addition, we crawled 3191 tweets on the 13 HK stocks in 2015 from Xueqiu, a Twitter-like investor social network in China. As there are more discussions on HK stocks in Xueqiu than in Guba, we collect the sentiment information related to HK stocks from Xueqiu rather than Guba. 

In the experiments, we use the data in the first 9 months as the training set and the last 3 months as the testing set. For the A-share dataset, 47.1\% of all the samples present upward trend, while 52.3\% present downward trend and 0.6\% keep still. For the HK stock market data, 54.9\% of all the samples present upward trend, while 42.3\% present downward trend and 2.8\% keep still. To remove the obstacles and obtain deterministic price moving trends, we set the movement scope threshold as 2\%. In particular, when the price change ratio of a stock is larger than 2\% (or smaller than -2\%) on a trading day, its price movement direction is considered as up (or down). Otherwise, this sample is considered as little fluctuation (or still) and excluded from our experiments. Therefore, our prediction task can be considered as a binary classification task, and can be addressed by a binary classifier.

%

\subsection{Comparison methods and metrics}

The following baselines and variations of our proposed model are implemented for comparison.

\begin{itemize}
\item \textbf{SVM}: we directly concatenate the stock quantitative features, event mode features and sentiment mode features as a linear vector, and use them as the input of SVM for prediction.

\item \textbf{PCA+SVM}: the Principle Component Analysis (PCA) technique is applied to reduce the dimensions of the original concatenated vector, and then the new vector is used as the input of SVM.

\item \textbf{TeSIA:} the tensor-based learning approach proposed in~\cite{li2015tensor} is the state-of-art baseline which utilizes multi-source information. Specifically, it uses a third-order tensor to model the firm-mode, event-mode and sentiment-mode data. Note that they construct an independent tensor for every stock on each trading day, without considering the correlations between stocks.

\item \textbf{CMT:} CMT denotes our proposed model, i.e., two auxiliary matrices and a tensor are factorized together. Note that the stock correlation matrices can be obtained by four methods. The default CMT is using the couple stock correlation method.  We will also evaluate CMT approaches with other stock correlation matrices as well, including co-evolving direction correlation, co-evolving p-change correlation and user perceived correlation.

\item \textbf{CMT-$\bm{Z}$:} To study the effectiveness of the stock correlation matrix $\bm{Z}$, we use the CMT without $\bm{Z}$ as a baseline.

\item \textbf{CMT-$\bm{Z}$-$\bm{X}$:} To study the effectiveness of the two auxiliary matrices $\bm{X}$ and $\bm{Z}$, we use only the tensor decomposition as a baseline.

\end{itemize}

Following the previous studies~\cite{ding2015deep,xie2013semantic}, the standard measures of accuracy (ACC) and Matthews Correlation Coefficient (MCC) are used as evaluation metrics. Larger values of the two metrics mean better performance.

\subsection{Prediction results}


\begin{table}
\renewcommand{\arraystretch}{1.3}
\small
\centering
\caption{ Prediction results of movement direction } \label{result1}
\vspace{2pt}
\begin{tabular}{ccccc} \hline
\multirow{2}{*}{Method} &
\multicolumn{2}{c}{China A-share} &
\multicolumn{2}{c}{HK} \\
  & ACC & MCC &ACC &MCC \\ \hline
SVM &55.37\% & 0.014 &55.13\% & 0.08\\ 
PCA+SVM &57.50\% &0.104 &56.07\% & 0.092\\ 
TeSIA  &60.63\% & 0.190 &60.38\% & 0.205\\ 
CMT-$\bm{Z}$-$\bm{X}$ &59.03\% &0.162  &59.36\% & 0.137\\ 
CMT-$\bm{Z}$  &60.25\% & 0.306 &60.29\% & 0.252\\ 
CMT &62.50\% & 0.409 &61.73\% & 0.331\\ \hline
\end{tabular}
\end{table}

Table~\ref{result1} shows the average stock movement prediction accuracy for all the 78 stocks  and 13 HK market stocks during the testing period. For the China stock market, it can be observed that our proposed CMT achieves the best performance, with 62.5\% in ACC and 0.409 in MCC. SVM method shows the worst performance, indicating that only using linear combination of the features cannot capture the coupling effects, which are crucial for improving the performance. PCA+SVM performs better than SVM, which is mainly because some noise can be removed by PCA. CMT outperforms CMT-$\bm{Z}$, indicating the stock correlation information plays an important role. Analogously, CMT-$\bm{Z}$ achieves better performance than CMT-$\bm{Z}$-$\bm{X}$, validating the effectiveness of the stock quantitative feature matrix. Therefore, both of the two auxiliary matrices provide effective knowledge to assist stock prediction. It can also be observed that compared to TeSIA, CMT improves accuracy by 3\% and significantly improves MCC by 115\%. We also evaluate our proposal on the HK stock market dataset, and the prediction results are similar as the A-share dataset. It demonstrates that the multi-task learning idea can be successfully applied to the stock prediction problem, resulting in a significant improvement in the prediction accuracy.

In addition to the superior prediction accuracy over TeSIA, our method also requires much fewer parameters to tune than TeSIA. Note that TeSIA ignores the relations among the stocks and takes each stock's prediction as a single-task regression learning. Due to the different characteristics of different stocks, learning a common regression function for all the stocks may not achieve the best performance. Thus, it requires TeSIA to learn different regression functions for different stocks respectively. Consequently, the number of parameters required to tune in TeSIA is linearly related to the amount of stocks, which is prohibitive due to the large volume of stocks in market.


\subsection{Results with different stock correlation methods}


\begin{table}
\renewcommand{\arraystretch}{1.3}
\small
\centering
\caption{ Results with different stock correlation methods } \label{result2}
\vspace{2pt}
\begin{tabular}{ccccc} \hline
\multirow{2}{*}{Method} &
\multicolumn{2}{c}{China A-share} &
\multicolumn{2}{c}{HK} \\
  & ACC & MCC & ACC &MCC \\ \hline
Without stock correlation matrix & 60.25\%& 0.306 &60.29\% & 0.252\\ 
Co-evolving direction correlation & 62.06\%& 0.382&61.39\% & 0.290\\ 
Co-evolving p-change correlation & 60.91\% & 0.300 &61.22\% & 0.205\\ 
User perceived correlation &61.64\% & 0.402 &60.77\% & 0.263\\ 
Coupled stock correlation (CMT) &62.50\%&0.409 &61.73\% & 0.331\\ \hline
\end{tabular}
\end{table}

As the stock correlation matrix can be obtained with different methods, we compare their performance in our framework and the results are shown in Table~\ref{result2}. It can be observed that for both of the two datasets, the coupled stock correlation method outperforms all the other methods in terms of both ACC and MCC. It implies that considering the coupled effects between attributes does help in getting more realistic correlations between stocks, and eventually results in better prediction capability. As a comparison, for the China A-share stocks, the performance with user perceived correlation takes the second place in terms of MCC and third place in terms of ACC, indicating that such information obtained from the social network can be used to capture the effective correlations as well. It makes sense as the co-occurrence relationships between stocks in users' postings are actually the results of the fundamental analysis to some extent, and represents the collective intelligence of the retail investors in the stock market. Moreover, this correlation with contexts in social media is more interpretable than the other quantitative methods. However, using user perceived correlations with HK stock data is not as good as that with China A-share data. The possible reason is that the number of HK stocks we choose and the amount of social media data related to HK stocks are less than those on China A-share stocks, and thus the obtained correlations for HK stocks may not be as accurate as the A-share stocks. The performance with co-evolving direction correlation performs better than with co-evolving p-change correlation for both of the two datasets. The difference between them is that p-change correlation takes the fluctuation scope of price into account as well. Thus, it indicates that movement direction is more important than movement scope for effectively mining the correlations. When compared to the approach without stock correlation matrix as an input, almost all the methods with stock correlations shows better performance, validating the effectiveness of this knowledge in stock prediction.

\subsection{Parameter sensitivity}

\begin{figure}
\setlength{\abovedisplayshortskip}{0pt}
\begin{center}
\begin{tabular}{c c c}
 \subfigure{\includegraphics[width=0.31\linewidth]{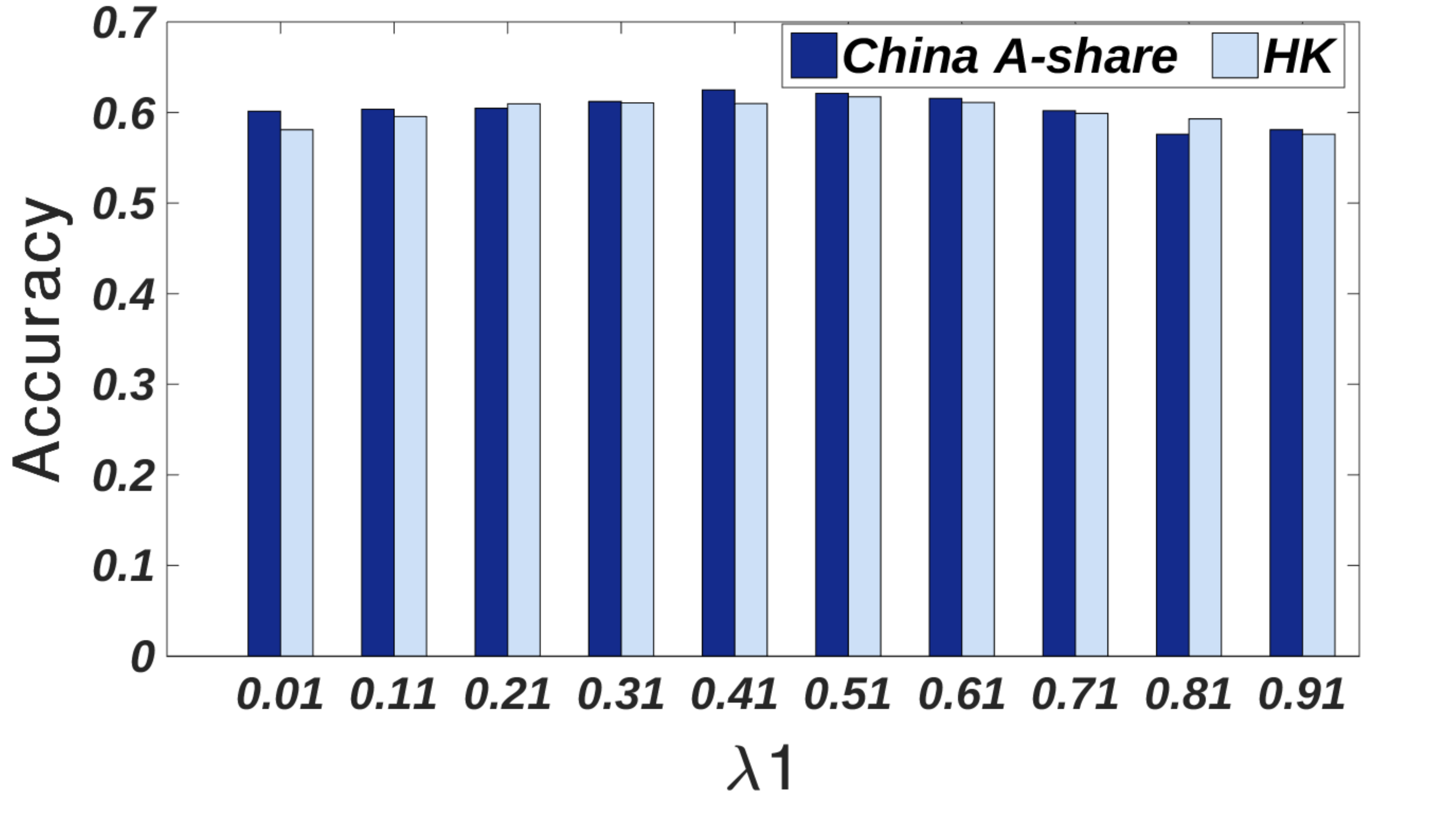}} & \subfigure{\includegraphics[width=0.31\linewidth]{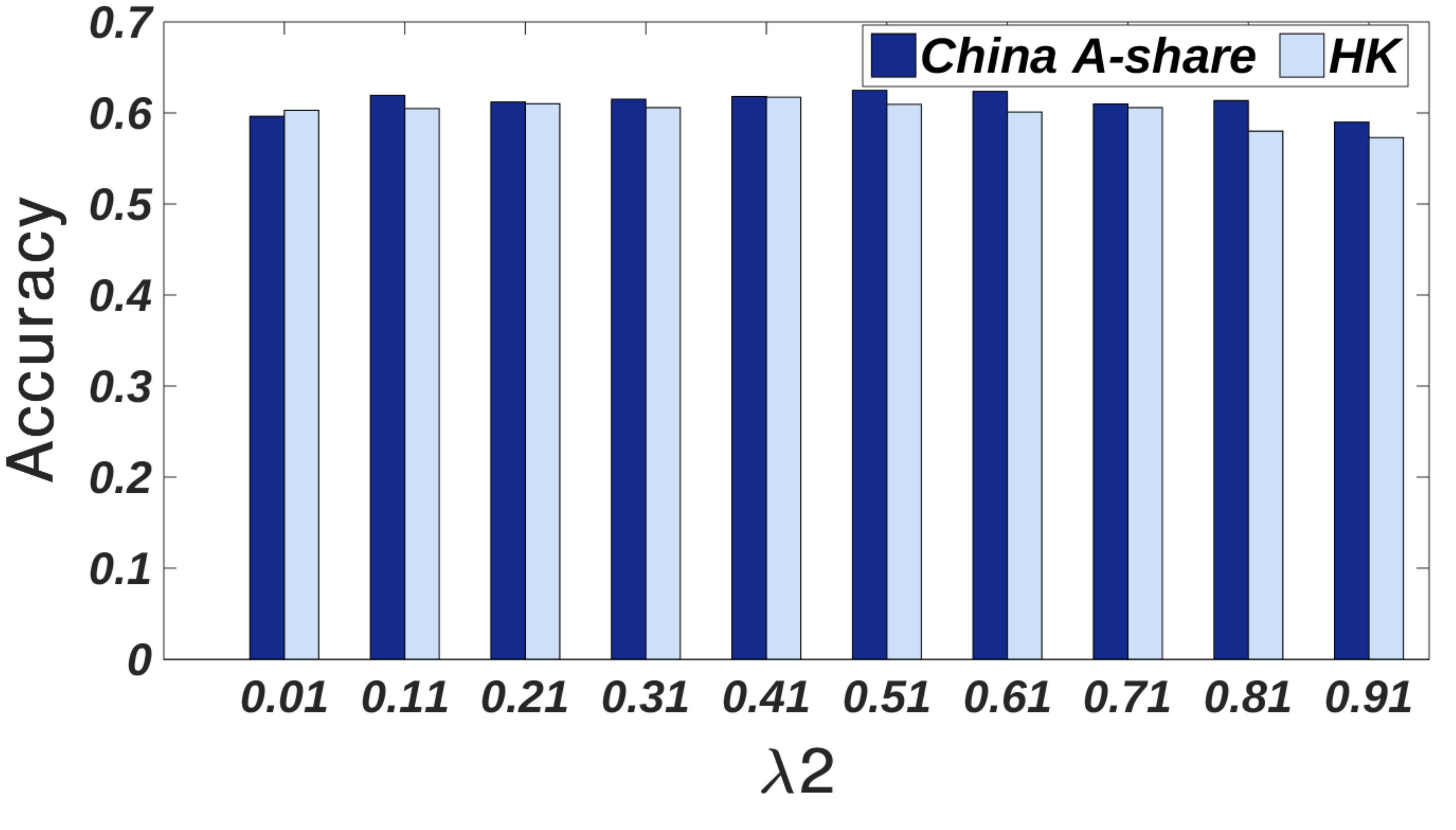}} & \subfigure{\includegraphics[width=0.31\linewidth]{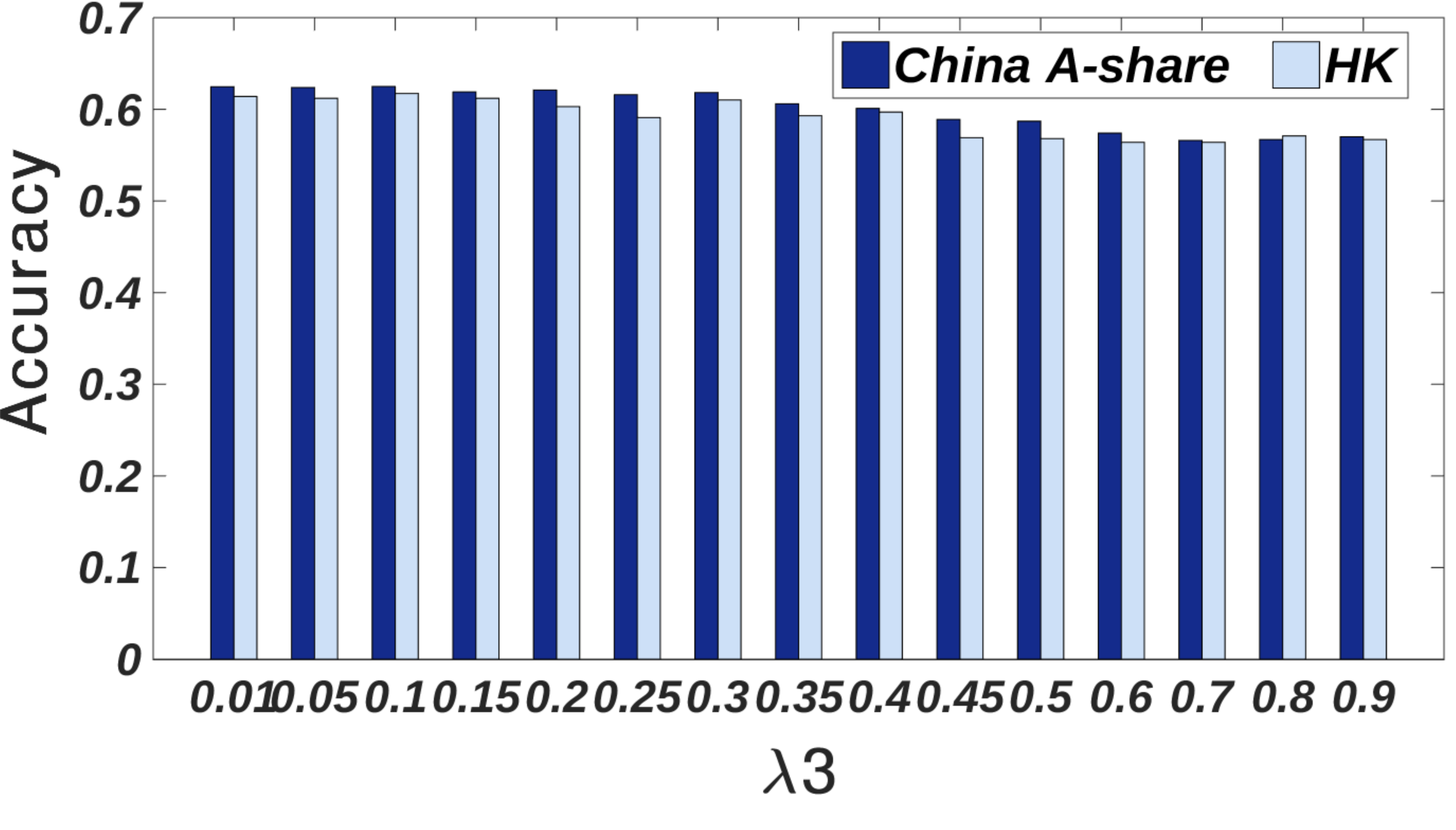}}\\
 (a)  ${\lambda_1}$ & (b)  ${\lambda_2}$ & (c)  ${\lambda_3}$\\
 \end{tabular}
      \caption{Sensitivity analysis on ${\lambda_1}$,  ${\lambda_2}$ and ${\lambda_3}$. x-axis represents the various values for the parameter and y-axis denotes the test accuracy.}
      \label{fig:parameter}
\end{center}
\end{figure}

%
%

There are three main parameters ${\lambda_1}$,  ${\lambda_2}$ and ${\lambda_3}$ in the proposed model Eq.(2). Next we conduct experiment to study the effects of the parameters on the model performance. To find the best setting of the parameters, we first fix the value of two parameters and then tune the value of the third parameter. With the discovered best setting of one parameter, we next tune another parameter in the same way. After several iterations, we can find the best parameter settings. Fig.~\ref{fig:parameter}(a),~\ref{fig:parameter}(b) and~\ref{fig:parameter}(c) show the effects of the three parameters on the model performance, respectively. For each figure, we fix the best settings of two parameters, and tune the value of the third one. It can be observed that the accuracy is relatively stable for both of the two datasets when ${\lambda_1}$ is within [0.21, 0.71],  ${\lambda_2}$ is within [0.21, 0.71] and ${\lambda_3}$ is within [0.01, 0.4].

\subsection{Case study on the stock correlation matrix}

\begin{table}
\centering
\caption{The stocks with top 10 highest correlation values of Sany Heavy Industry} \label{highestSany}
\begin{adjustbox}{max width=\columnwidth}
\begin{tabular}{ c c c c c} \hline
Rank & Code & Name &Industry & Correlation\\ \hline
1 &600150.SH &China CSSC Co., Ltd. &Machinery &0.806\\
2 &601989.SH &China Shipbuilding Industry Co., Ltd.&Machinery &0.800 \\ 
3 &601766.SH &CRRC Co., Ltd.&Machinery &0.773 \\ 
4 &601668.SH &China State Construction Engineering Co., Ltd. &Construction &0.659\\ 
5 &601618.SH &Metallurgical Corporation of China Ltd. &Construction &0.616\\ 
6 &601600.SH &Aluminum Corporation of China Limited &Material &0.615\\ 
7 &600010.SH &Inner Mongolia Baotou Steel Union Co., Ltd.&Material &0.613\\ 
8 &601088.SH &China Shenhua Energy Co., Ltd.&Energy &0.612\\ 
9 &601699.SH &Power Construction Co., Ltd.&Construction &0.608\\ 
10&601186.SH &China Railway Construction Co., Ltd.&Construction &0.607\\ \hline
\end{tabular}
\end{adjustbox}
\end{table}

\begin{table}
\small
\centering
\caption{The stocks with top 5 lowest correlation values of Sany Heavy Industry} \label{lowestSany}
\begin{adjustbox}{max width=\columnwidth}
\begin{tabular}{ccccc} \hline
Rank & Code & Name & Industry & Correlation\\ \hline
1 &600518.SH &Kangmei Pharmaceutical Co., Ltd. &Medical &0.183\\ 
2 &300104.SZ &Leshi Internet Information \& Technology Corp., BeiJing	&Internet &0.200 \\ 
3 &600637.SH &Shanghai Oriental Pearl Media Co., Ltd.&Media &0.202 \\ 
4 &601727.SH &Shanghai Electric Group Co., Ltd. &Appliance &0.227\\ 
5 &600050.SH &China United Network Communications Co., Ltd. &Telecommunications &0.233\\  \hline
\end{tabular}
\end{adjustbox}
\end{table}

\begin{table}
\small
\centering
\caption{The stocks with top 10 highest correlation values of China Railway Construction} \label{highestRailway}
\begin{adjustbox}{max width=\columnwidth}
\begin{tabular}{ccccc} \hline
Rank & Code & Name & Industry & Correlation\\ \hline
1 &601390.SH &China Railway Group Limited &Construction &0.894\\ 
2 &601800.SH &China Communications Construction Co., Ltd.	&Construction &0.869 \\ 
3 &601668.SH &China State Construction Engineering Co., Ltd.&Construction &0.862 \\ 
4 &601699.SH &Power Construction Co., Ltd. &Construction &0.852\\ 
5 &601618.SH &Metallurgical Corporation of China Ltd. &Construction &0.831\\ 
6 &601818.SH &China Everbright Bank Co., Ltd. &Bank &0.602\\ 
7 &000001.SZ &Ping An Bank Co., Ltd.	&Bank &0.583\\ 
8 &601766.SH &CRRC Co., Ltd.&Machinery &0.569\\
9 &600015.SH &Hua Xia Bank Co., Ltd.&Bank &0.563\\ 
10&600031.SH &Sany Heavy Industry Co., Ltd.&Machinery &0.554\\ \hline
\end{tabular}
\end{adjustbox}
\end{table}

\begin{table}
\small
\centering
\caption{The stocks with top 5 lowest correlation values of China Railway Construction} \label{lowestRailway}
\begin{adjustbox}{max width=\columnwidth}
\begin{tabular}{ccccc} \hline
Rank & Code & Name & Industry & Correlation\\ \hline
1 &600518.SH &Kangmei Pharmaceutical Co., Ltd. &Medical &0.112\\ 
2 &600893.SH &AVIC Aviation Engine Corporation PLC.&Aerospace &0.144 \\ 
3 &600637.SH &Shanghai Oriental Pearl Media Co., Ltd.&Media &0.146 \\ 
4 &002024.SZ &Suning Commerce Group Co., Ltd. &Elec retail &0.160\\ 
5 &601727.SH &Shanghai Electric Group Co., Ltd. &Appliance &0.189\\ \hline
\end{tabular}
\end{adjustbox}
\end{table}

To demonstrate the effectiveness and interpretability of the stock correlation results, we select two stocks, Sany Heavy Industry (stock code 600031) and China Railway Construction (stock code 601186), from A-share stocks as two examples to show their correlations with other stocks. Specifically, using the coupled stock similarity method, the top 10 most correlated stocks with Sany Heavy Industry are listed in Table~\ref{highestSany}. It can be observed that the most relevant stocks (top 3) belong to the same industry category (i.e., Machinery) as Sany, with correlation values larger than 0.7\%. The other correlated stocks do not belong to the same industry, but they are correlated to the upstream or downstream of the Machinery Industry, with the correlation values ranging from 0.6 to 0.7. Table~\ref{lowestSany} shows the top 5 lowest correlated stocks with Sany, which are all unrelated stocks with correlation value lower than 0.2. Similar observations can also be obtained for the correlation results with China Railway Construction shown in Table~\ref{highestRailway} and Table~\ref{lowestRailway}. Hence, the correlation values exhibit good interpretability, and do play significant roles in our framework.
 
%
%
%
%

\section{Conclusions and future work}

In this paper, we have proposed a coupled matrix and tensor factorization method for stock market prediction. Different from previous studies that commonly exploit only one or two data sources,  our model integrates events, sentiments and quantitative features extracted from various data sources, involving the Web news, social media and the stock quantitative data provider. A stock movement tensor is then constructed to model the intrinsic relations among stocks, events and sentiments. However, it is challenging to decompose the tensor due to the sparsity of the events. To address this issue, we learn the stock correlations with various methods, and apply this knowledge to facilitate the multi-task learning by collaboratively factorizing the tensor with two auxiliary matrices. With the factorized low rank matrices, we can effectively predict the stock movements by completing the missing values in the sparse tensor. Evaluations on the China A-share stock data and the HK stock data show the effectiveness of our model.

Potential directions of future works include incorporating other data sources, e.g., knowledge graph information, to enrich the sparse events. Advanced NLP techniques, e.g., learning event representations with domain knowledge, are also required for better events categorization. In addition, it is challenging to propose a domain-specific sentiment classification approach with better accuracy. It would be interesting to investigate how to extend our model to a time series prediction framework. According to existing studies in behavioral finance, the decision-making process in financial market is affected by psychological biases such as overconfidence, which can also also be taken into account for prediction in future work.

\section{Acknowledgements} 

This work has been supported in part by the State Key Development Program of Basic Research of China (No. 2013CB329604), the Natural Science Foundation of China (No. 61300014, 61370068, 61602237), DongGuan Innovative Research Team Program (No. 201636000100038), and NSF through grants IIS-1526499, and CNS-1626432.



\section*{References}

\begin{thebibliography}{}

\bibitem{fama1965behavior}
E.~F. Fama, The behavior of stock-market prices, The journal of Business 38~(1)
  (1965) 34--105.

\bibitem{PAN201790}
Y.~Pan, Z.~Xiao, X.~Wang, D.~Yang, A multiple support vector machine approach
  to stock index forecasting with mixed frequency sampling, Knowledge-Based
  Systems 122 (2017) 90 -- 102.

\bibitem{whitehouse}
S.~C. Johnson, Analysis: False white house tweet exposes instant trading
  dangers,
  \url{http://www.reuters.com/article/us-usa-markets-tweet-idUSBRE93M1FD20130423},
  2013, {(}accessed 17.01.14{)}.

\bibitem{ding2014using}
X.~Ding, Y.~Zhang, T.~Liu, J.~Duan, Using structured events to predict stock
  price movement: An empirical investigation, in: The Conference on Empirical
  Methods on Natural Language Processing (EMNLP-14), 2014, pp. 1415--1425.

\bibitem{ding2015deep}
X.~Ding, Y.~Zhang, T.~Liu, J.~Duan, Deep learning for event-driven stock
  prediction, in: Proceedings of the 24th International Joint Conference on
  Artificial Intelligence (ICJAI-15), 2015, pp. 2327--2333.

\bibitem{dingknowledge}
X.~Ding, Y.~Zhang, T.~Liu, J.~Duan, Knowledge-driven event embedding for stock
  prediction, in: Proceedings of the 26th International Conference on
  Computational Linguistics (COLING-16), 2016, pp. 2133--2142.

\bibitem{Prechter1999}
R.~R. Prechter, The wave principle of human social behavior and the new science
  of socionomics, Vol.~1, New Classics Library, 1999.

\bibitem{Nofsinger2005}
J.~R. Nofsinger, Social mood and financial economics, J. Finance 6~(3) (2005)
  144--160.

\bibitem{bollen2011twitter}
J.~Bollen, H.~Mao, X.~Zeng, Twitter mood predicts the stock market, J. Comput.
  Sci. 2~(1) (2011) 1--8.

\bibitem{nguyen2015topic}
T.~H. Nguyen, K.~Shirai, Topic modeling based sentiment analysis on social
  media for stock market prediction, in: Proceedings of the 53rd Annural
  Meeting of the Association for Computational Linguistics (ACL-15), 2015.

\bibitem{feldman2011stock}
R.~Feldman, B.~Rosenfeld, R.~Bar-Haim, M.~Fresko, The stock sonar-sentiment
  analysis of stocks based on a hybrid approach, in: Twenty-Third IAAI
  Conference, 2011.

\bibitem{li2015tensor}
Q.~Li, L.~Jiang, P.~Li, H.~Chen, Tensor-based learning for predicting stock
  movements, in: The Twenty-Ninth AAAI Conference on Artificial Intelligence
  (AAAI-15), 2015, pp. 1784--1790.

\bibitem{li2016tensor}
Q.~Li, Y.~Chen, L.~L. Jiang, P.~Li, H.~Chen, A tensor-based information
  framework for predicting the stock market, ACM Trans. Inform. Syst. (TOIS)
  34~(2) (2016) 11.

\bibitem{cutler1988moves}
D.~M. Cutler, J.~M. Poterba, L.~H. Summers, What moves stock prices, J. Portf.
  Manag. 15 (1989) 4--12.

\bibitem{tetlock2008more}
P.~C. Tetlock, M.~SAAR-TSECHANSKY, S.~Macskassy, More than words: Quantifying
  language to measure firms' fundamentals, J. Finance 63~(3) (2008) 1437--1467.

\bibitem{luss2015predicting}
R.~Luss, A.~D'Aspremont, Predicting abnormal returns from news using text
  classification, Quantitative Finance 15~(6) (2015) 999--1012.

\bibitem{xie2013semantic}
B.~Xie, R.~J. Passonneau, L.~Wu, G.~G. Creamer, Semantic frames to predict
  stock price movement, in: Proceedings of the 51st annual meeting of the
  association for computational linguistics, 2013, pp. 873--883.

\bibitem{wang2014semiparametric}
W.~Y. Wang, Z.~Hua, A semiparametric gaussian copula regression model for
  predicting financial risks from earnings calls, in: The 51st Annual Meeting
  of the Association for Computational Linguistics (ACL-14), 2014, pp.
  1155--1165.

\bibitem{peng2015leverage}
Y.~Peng, H.~Jiang, Leverage financial news to predict stock price movements
  using word embeddings and deep neural networks, arXiv preprint
  arXiv:1506.07220.

\bibitem{kumar2016survey}
B.~S. Kumar, V.~Ravi, A survey of the applications of text mining in financial
  domain, Knowl.-Based Syst. 114 (2016) 128--147.

\bibitem{changmeasuring}
C.-Y. Chang, Y.~Zhang, Z.~Teng, Z.~Bozanic, B.~Ke, Measuring the information
  content of financial news, in: Proceedings of the 26th International
  Conference on Computational Linguistics (COLING'16), 2016, pp. 3216--3225.

\bibitem{yoshihara2014predicting}
A.~Yoshihara, K.~Fujikawa, K.~Seki, K.~Uehara, Predicting stock market trends
  by recurrent deep neural networks, in: Pacific Rim International Conference
  on Artificial Intelligence, Springer, 2014, pp. 759--769.

\bibitem{schumaker2009quantitative}
R.~P. Schumaker, H.~Chen, A quantitative stock prediction system based on
  financial news, Information Processing \& Management 45~(5) (2009) 571--583.

\bibitem{schumaker2009textual}
R.~P. Schumaker, H.~Chen, Textual analysis of stock market prediction using
  breaking financial news: The {AZFin} text system, ACM Trans. Inform. Syst.
  (TOIS) 27~(2) (2009) 1139--1141.

\bibitem{li2014news}
X.~Li, H.~Xie, L.~Chen, J.~Wang, X.~Deng, News impact on stock price return via
  sentiment analysis, Knowl.-Based Syst. 69 (2014) 14--23.

\bibitem{wang2012novel}
B.~Wang, H.~Huang, X.~Wang, A novel text mining approach to financial time
  series forecasting, Neurocomputing 83 (2012) 136--145.

\bibitem{schumaker2012evaluating}
R.~P. Schumaker, Y.~Zhang, C.-N. Huang, H.~Chen, Evaluating sentiment in
  financial news articles, Decision Support Systems 53~(3) (2012) 458--464.

\bibitem{liang2013associating}
X.~Liang, R.-C. Chen, Y.~He, Y.~Chen, Associating stock prices with web
  financial information time series based on support vector regression,
  Neurocomputing 115 (2013) 142--149.

\bibitem{li2014media}
Q.~Li, T.~Wang, Q.~Gong, Y.~Chen, Z.~Lin, S.-k. Song, Media-aware quantitative
  trading based on public web information, Decision support systems 61 (2014)
  93--105.

\bibitem{zhang2011predicting}
X.~Zhang, H.~Fuehres, P.~A. Gloor, Predicting stock market indicators through
  twitter ?{I} hope it is not as bad as {I} fear, Proc.-Soc. Behav. Sci. 26
  (2011) 55--62.

\bibitem{si2013exploiting}
J.~Si, A.~Mukherjee, B.~Liu, Q.~Li, H.~Li, X.~Deng, Exploiting topic based
  twitter sentiment for stock prediction, in: The 51st Annual Meeting of the
  Association for Computational Linguistics (ACL-13), 2013, pp. 24--29.

\bibitem{si2014exploiting}
J.~Si, A.~Mukherjee, B.~Liu, S.~J. Pan, Q.~Li, H.~Li, Exploiting social
  relations and sentiment for stock prediction, in: The Conference on Empirical
  Methods on Natural Language Processing (EMNLP-14), Vol.~14, 2014, pp.
  1139--1145.

\bibitem{li2014effect}
Q.~Li, T.~Wang, P.~Li, L.~Liu, Q.~Gong, Y.~Chen, The effect of news and public
  mood on stock movements, Inform. Sci. 278 (2014) 826--840.

\bibitem{ruiz2012correlating}
E.~J. Ruiz, V.~Hristidis, C.~Castillo, A.~Gionis, A.~Jaimes, Correlating
  financial time series with micro-blogging activity, in: Proceedings of the
  fifth ACM international conference on Web search and data mining, ACM, 2012,
  pp. 513--522.

\bibitem{DUAN2013151}
J.~Duan, H.~Liu, J.~Zeng, Posterior probability model for stock return
  prediction based on analyst?s recommendation behavior, Knowledge-Based
  Systems 50 (2013) 151 -- 158.

\bibitem{behaviorbook}
L.~Ackert, R.~Deaves, Behavioral Finance: Psychology, Decision-Making, and
  Markets, South-Western: Cengage Learning, 2010.

\bibitem{maplosone}
M.~A. Bertella, F.~R. Pires, L.~Feng, H.~E. Stanley, Confidence and the stock
  market: An agent-based approach, PLOS ONE 9~(1) (2014) 1--9.

\bibitem{Chuang2006An}
W.~I. Chuang, B.~S. Lee, An empirical evaluation of the overconfidence
  hypothesis, Journal of Banking \& Finance 30~(9) (2006) 2489--2515.

\bibitem{PhysRevE.70.026101}
W.-J. Ma, C.-K. Hu, R.~E. Amritkar, Stochastic dynamical model for stock-stock
  correlations, Phys. Rev. E 70 (2004) 026101.

\bibitem{Chen938934}
P.~Chen, Modelling the stochastic correlation, Master's thesis, KTH,
  Mathematical Statistics (2016).

\bibitem{sepp2011modeling}
S.~Sepp, Modeling of stock return correlation, Ph.D. thesis, Master thesis,
  Universiteit van Amsterdam (2011).

\bibitem{preis2012quantifying}
T.~Preis, D.~Y. Kenett, H.~E. Stanley, D.~Helbing, E.~Ben-Jacob, Quantifying
  the behavior of stock correlations under market stress, Scientific reports 2
  (2012) 752.

\bibitem{wang2011coupled}
C.~Wang, L.~Cao, M.~Wang, J.~Li, W.~Wei, Y.~Ou, Coupled nominal similarity in
  unsupervised learning, in: Proceedings of the 20th ACM international
  conference on Information and knowledge management (CIKM-11), ACM, 2011, pp.
  973--978.

\bibitem{Li2015}
F.~Li, G.~Xu, L.~Cao, Coupled matrix factorization within non-iid context, in:
  Pacific-Asia Conference on Knowledge Discovery and Data Mining, Springer,
  2015, pp. 707--719.

\bibitem{fama1993common}
E.~F. Fama, K.~R. French, Common risk factors in the returns on stocks and
  bonds, Journal of financial economics 33~(1) (1993) 3--56.

\bibitem{hownet}
Z.~Dong, Hownet knowledge database, \url{http://www.keenage.com/}, 2011,
  {(}accessed 17.01.15{)}.

\bibitem{mikolov2013distributed}
T.~Mikolov, I.~Sutskever, K.~Chen, G.~S. Corrado, J.~Dean, Distributed
  representations of words and phrases and their compositionality, in: Advances
  in neural information processing systems, 2013, pp. 3111--3119.

\bibitem{zhang2016effective}
X.~Zhang, Y.~Yao, Y.~Ji, B.~Fang, Effective and fast near duplicate detection
  via signature-based compression metrics, Mathematical Problems in Engineering
  2016.

\bibitem{ku2007mining}
L.-W. Ku, H.-H. Chen, Mining opinions from the web: Beyond relevance retrieval,
  J. American Soc. Inf. Science Tech. 58~(12) (2007) 1838--1850.

\bibitem{karatzoglou2010multiverse}
A.~Karatzoglou, X.~Amatriain, L.~Baltrunas, N.~Oliver, Multiverse
  recommendation: n-dimensional tensor factorization for context-aware
  collaborative filtering, in: Proceedings of the fourth ACM conference on
  Recommender systems, ACM, 2010, pp. 79--86.

\bibitem{wang2015citywide}
S.~Wang, L.~He, L.~Stenneth, P.~S. Yu, Z.~Li, Citywide traffic congestion
  estimation with social media, in: Proceedings of the 23rd SIGSPATIAL
  International Conference on Advances in Geographic Information Systems, ACM,
  2015, p.~34.

\bibitem{webnews}
X.~Zhang, Z.~Yunjia, Financial web news dataset,
  \url{https://pan.baidu.com/s/1mhCLJJi}, 2017, {(}accessed 17.01.15{)}.

\bibitem{gubadata}
X.~Zhang, Y.~Yao, Guba dataset, \url{https://pan.baidu.com/s/1i5zAWh3}, 2017,
  {(}accessed 17.01.15{)}.

\end{thebibliography}
\end{document}